\documentclass{jpsj-suppl}

\title{Low-energy Antikaon Interaction with Nuclei: The AMADEUS Challenge}

\author{J. \textsc{Marton}$^{1}$, M. \textsc{Bazzi}$^{2}$, G. \textsc{Bellotti}$^{3,4}$, C. \textsc{Berucci}$^{1}$, D. \textsc{Bosnar}$^{5}$, A.M. \textsc{Bragadireanu}$^{6}$,
C. \textsc{Curceanu}$^{2}$, A. \textsc{Clozza}$^{2}$, M. \textsc{Cargnelli}$^{1}$, A.D. \textsc{Butt}$^{3,4}$, R. \textsc{Del Grande}$^{2}$, L. \textsc{Fabbietti}$^{7}$, C. \textsc{Fiorini}$^{3,4}$,
F. \textsc{Ghio}$^{2}$, C. \textsc{Guaraldo}$^{2}$, M. \textsc{Iliescu}$^{2}$, P. \textsc{Levi Sandri}$^{2}$,  D. \textsc{Pietreanu}$^{6}$,
K. \textsc{Piscicchia}$^{2,8}$, A. \textsc{Romero Vidal}$^{9}$, A. \textsc{Scordo}$^{2}$, H. \textsc{Shi}$^{2}$, D. \textsc{Sirghi}$^{2,6}$, F. \textsc{Sirghi}$^{2,6}$,
I. \textsc{Tucakovic}$^{10}$, O. \textsc{Vazquez Doce}$^{7}$, E. \textsc{Widmann}$^{1}$ and J. \textsc{Zmeskal}$^{1}$}

\inst{
$^{1}$ Stefan-Meyer-Institut f\"{u}r subatomare Physik, Vienna, Austria \\
$^{2}$ INFN, Laboratori Nazionali di Frascati, CP 13, Via E. Fermi 40, I-00044, Frascati (Roma), Italy \\
$^{3}$ Politecnico di Milano, Dipartimento di Elettronica, Informazione e Bioingegneria, Piazza L. da Vinci 32, 20133, Milano, Italy   \\
$^{4}$ INFN Sezione di Milano, Via Celoria 16, 20133, Milano, Italy\\
$^{5}$ Physics Department, University of Zagreb, Zagreb, Croatia \\
$^{6}$ Horia Hulubei National Institute of Physics and Nuclear Engineering (IFIN-HH), Magurele,
Romania \\
$^{7}$ Excellence Cluster Universe, Technische Universit\"at M\"unchen, Garching, Germany \\
$^{8}$ Museo Storico della Fisca e Centro Studi e Ricerche ``Enrico Fermi'', Piazza del Viminale 1-00184 Roma, Italy\\
$^{9}$ Universidade de Santiago de Compostela, Santiago de Compostela, Spain \\
$^{10}$ Ruder Boskovic Institute, Beijenicka cesta 54, Zagreb, Croatia \\
}

\email{johann.marton@oeaw.ac.at}

\recdate{May 30, 2016}

\abst{
The low-energy strong interaction of antikaons (K$^{-}$) with nuclei has many facets and represents a lively and challenging research field. It is interconnected to the peculiar role of strangeness, since the strange quark is rather light, but still much heavier than the up and down quarks. Thus, when strangeness is involved one has to deal with spontaneous and explicit symmetry breaking in QCD. It is well known that the antikaon interaction with nucleons is attractive, but how strong ? Is the interaction strong enough to bind nucleons to form kaonic nuclei and, if so, what are the properties (binding energy, decay width)? There are controversial indications for such bound states and new results are expected to come soon.
The existence of antikaon mediated bound states might have important consequences since it would open the possibility for the formation of cold baryonic matter of high density which might have a severe impact in astrophysics for the understanding of the composition of compact (neutron) stars.
New experimental opportunities could be provided by the AMADEUS experiment at the DA$\Phi$NE electron-positron collider at LNF-INFN (Frascati, Italy). Pre-AMADEUS studies on the antikaon interaction with nuclei are carried out by analysis of data collected by KLOE in till 2005 and in special data runs using a carbon target insert. Studies for the dedicated AMADEUS detector setup taking advantage of the low-energy antikaons from $\Phi$-meson decay delivered by DA$\Phi$NE are in progress. Some results obtained so far and the perspectives of the AMADEUS experiment are presented and discussed.}

\kword{antikaon reactions, antikaon nuclear absorption, bound states, hyperon resonances}

\begin{document}
\maketitle

\section{Introduction}

%
%
%
In the last years significant progress has been made in the experimental studies of the antikaon (K$^{-}$) interaction on nucleons \cite{siddharta12} and in the correlated theory of strong interaction with strangeness \cite{weise15,kaiser95,oset98,oller01,ikeda12}. The SIDDHARTA experiment at the DA$\Phi$NE electron-positron collider obtained the most precise results on the antikaon-proton interaction via X-ray spectroscopy of kaonic hydrogen. On the other hand the strong interaction of K$^{-}$ with nuclei is investigated in several experiments currently, like experiments at J-PARC \cite{e15, e27}. Complementary, subthreshold resonances in the strangeness sector with hitherto unclear nature like $\Lambda$(1405)\cite{hyodo12} are studied with p+p reaction with HADES at GSI \cite{laura} and at JLab \cite{moriya13}. \\
However, best suited for low-energy strangeness nuclear physics is DA$\Phi$NE which is a unique source of nearly mono-energetic low-energy antikaons emitted with a back-to-back topology in the $\Phi$ decay. Therefore, the focus of studies on open problems is directed to KLOE with the extension to AMADEUS (see below).

\section{AMADEUS}

The antikaon interaction on nuclei resembles an interplay between explicit and spontaneous chiral symmetry breaking. From experimental studies like the results of SIDDHARTA and theoretical studies follows that the antikaon strong interaction at low energies is strongly attractive. Less clear is the quantitative strength of the interaction connected with the question of the existence of possible kaonic nuclear bound states, and if existent their production mechanism and properties. The rich physics program of AMADEUS (Antikaon Matter At DA$\Phi$NE: Experiments with Unravelling Spectroscopy)\cite{LOI2006,AMADEUS} covers the experimental research from antikaon(kaon) scattering at lowest momenta, kaonic nuclear bound state searches, studies of antikaon nuclear absorption processes and sub-threshold resonances like $\Lambda$(1405) in the s-wave and $\Sigma$(1385) in the p-wave as well as hypernuclear physics.

\subsection{Pre-AMADEUS Studies}
\begin{figure}[tbh]
\includegraphics [width= 90mm]{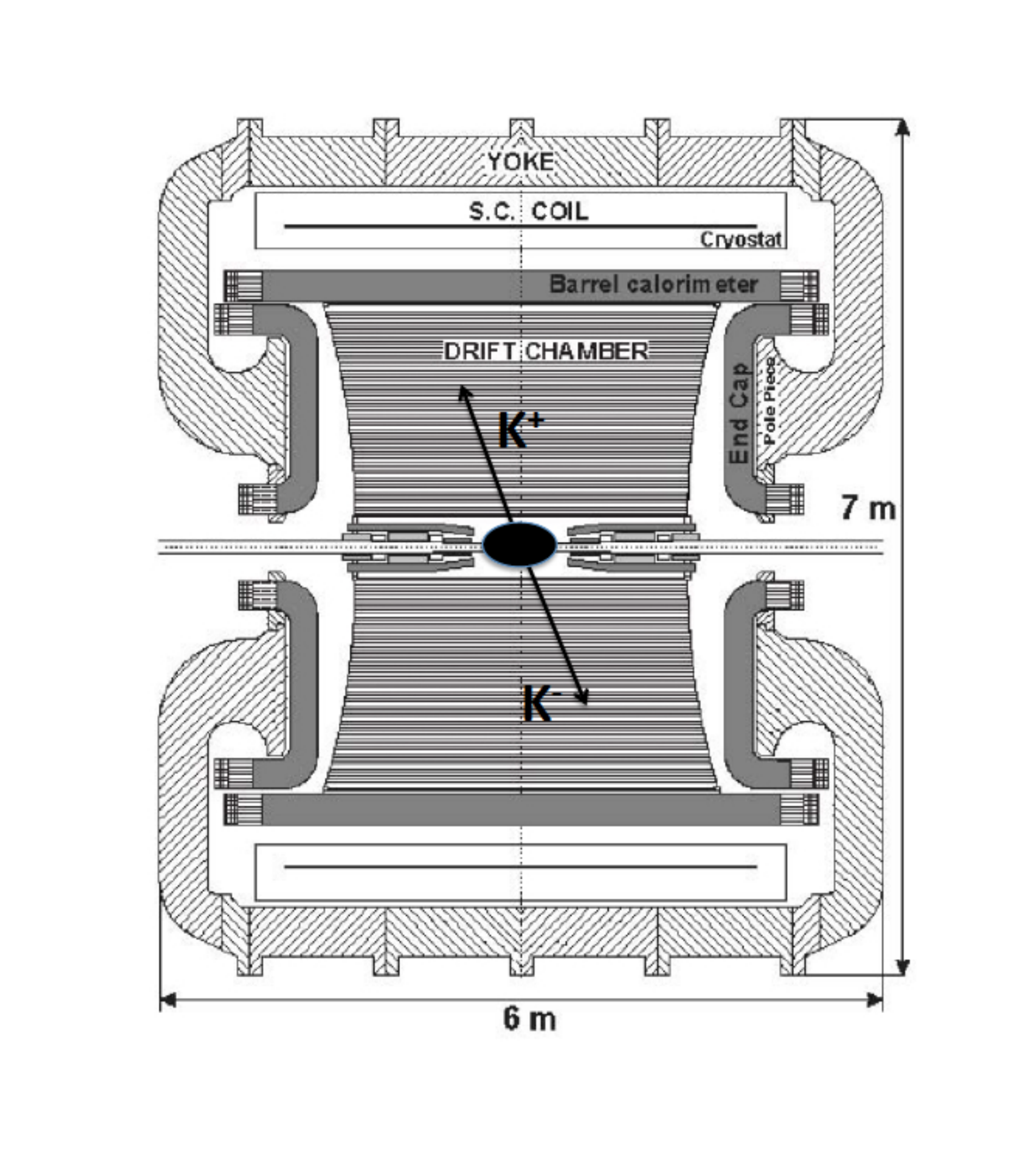}
\caption{Scheme of the KLOE detector consisting of magnetic yoke, electromagnetic calorimeter and the drift chamber surrounding the electron-positron intersection of DA$\Phi$NE. Antikaon induced reactions in the DC gas after $\Phi$ decay in K$^{+}$+K$^{-}$ can be studied with the KLOE detector.}
\label{f1}
\end{figure}

The KLOE detector has a 4$\pi$ geometry surrounding the e$^{+}$e$^{-}$ collision zone of DA$\Phi$NE. The detector system consists of a large cylindrical drift chamber (DC) \cite{adinolfi02} and a sampling electromagnetic calorimeter (EMC)\cite{adinolfi02a} inside a superconducting magnet (magnetic field $\sim$ 0.52T). The DC has a radius of 2m and a length of 3m and provides a spacial resolution of 150$\mu$m (radial) and 2mm (longitudinal). The EMC  has a nearly full 4$\pi$  solid angle providing a time resolution of $\sigma_{t}$= 54 ps/$\sqrt{E}$ and an energy resolution of $\sigma_{E}$/E=5.7\%/$\sqrt{E}$ (E given in GeV). It has to be highlighted that the EMC is capable of the detection of charged and neutral particles which is an important feature for the experimental studies. Overall the KLOE detector is a powerful device for the study of kaon physics at the $\Phi$ factory DA$\Phi$NE which delivers low-energy kaons (momentum $\sim$ 120 MeV/c) from the decay of the resonantly produced $\Phi$ vector mesons. The decay reaction

\begin{equation}\label{1}
  \Phi \rightarrow K^{+} + K^{-}
\end{equation}
has a branching ratio of about 50\% and the back-to-back emitted kaons are nearly mono-chromatic at low energy. Therefore, the K$^{-}$ emitted in the $\Phi$ decay are perfectly suited for investigations of kaon-induced reactions in low-density gases. The principal setup of KLOE with the decaying $\Phi$ meson in the e$^{+}$$e^{-}$ intersection region is displayed in Fig.\ref{f1}

The gas mixture of the KLOE central drift chamber consists of $^{4}$He (90\%) and isobutane C$_{4}$H$_{10}$ (10\%). A small fraction of K$^{-}$ ($\sim$0.1\%) is stopping in the gas and thus enables the antikaon absorption processes and the search for kaonic bound states like the prototype
system K$^{-}$pp proposed by theory \cite{AY2002}. Indications for the existence were found in different experiments like \cite{suzuki2012,agnello05,tsuzuki08} but there is no clear picture concerning binding energy and decay width up to now.

As a first step KLOE data (about 2fb$^{-1}$ collected in the period till 2005) are analyzed. In-flight reactions of K$^{-}$ on the different materials of the KLOE detector and the DC gas can be studied in invariant mass spectroscopy.\\
In the following recent selected studies using KLOE data are presented. More detailed information about the results of the Pre-AMADEUS studies can be found in refs. \cite{catalina14,Kristian,vazquez12 }.

\subsubsection{Studies on K$^{-}$ absorption processes and search for kaonic nuclear bound state}
\begin{figure}[tbh]
\includegraphics [width= 160mm]{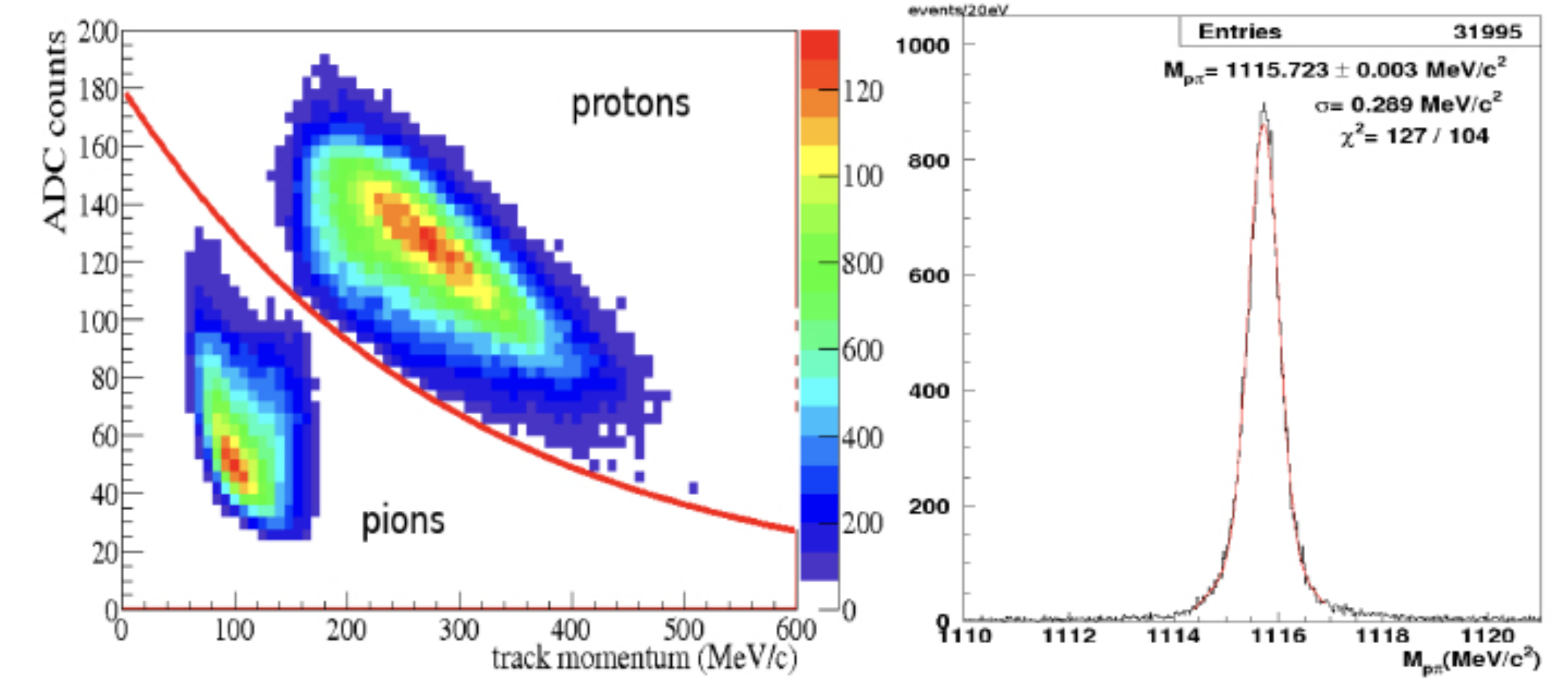}
\caption{Left: Clear separation of pions and protons as function of the track momentum. The red solid line gives the proton selection function. Right: Invariant mass spectrum of $\Lambda$(1116)for selected pairs of $\pi$-proton.}
\label{f2}
\end{figure}

The study  of the K$^{-}$ nucleon and K$^{-}$multi-nucleon absorption concerns also investigation of possible antikaon
multi-nucleon bound states properties. These studies proceed via the analysis of the $\Lambda$p, $\Sigma^{0}$p, $\Lambda$d (expected decay channels of eventual K$^{-}$pp and K$^{-}$ppn bound states) and $\Lambda$t correlations. The K$^{-}$ absorption in the $\Sigma^{0}$p final state yielded the first exclusive determination of the 2-nucleon absorption. It was performed with KLOE data with a total integrated luminosity of 1.74 fb$^{-1}$ collected 2004-2005 \cite{oton16}. The presence of a hyperon $\Lambda$(1116) always indicate a K$^{-}$ hadronic interaction with KLOE material. The first step for selecting the K$^{-}$ absorption reaction in $\Sigma^{0}$p is the identification of the hyperon $\Lambda$(1116) via its decay into proton and $\pi^{-}$. The clear separation of proton and pion and the reconstructed invariant mass spectrum of $\Lambda$(1116) is shown in fig. \ref{f2}. After reconstruction of the $\Lambda$(1116) vertex the tracks of other particles like pions are searched in the DC with their dE/dx.




Furthermore a possible bound state K$^{-}$pp was searched in the $\Sigma^{0}$p final state. A yield of 0.044 for K$^{-}$pp per stopped K$^{-}$ is extracted with low statistical significance (1 $\sigma$).

\subsubsection{Study of $\Lambda$(1404)}
The investigation of the resonance $\Lambda$(1405) is performed through its decay in $\Sigma$$\pi$. Here the neutral decay of $\Lambda$(1405)

\begin{equation}\label{e2}
  \Lambda(1405) \rightarrow \Sigma^{0} + \pi^{0}
\end{equation}

has many advantages for the analysis (golden channel): The $\Sigma^{0}$ decays

\begin{equation}\label{3}
  \Sigma^{0} \rightarrow \Lambda + \gamma
\end{equation}
and the $\pi^{0}$ decay into

\begin{equation}\label{4}
  \pi^{0} \rightarrow \gamma + \gamma
\end{equation}

Therefore one has in the final state (p$\pi^{-}$ and 3$\gamma$ which helps to discriminate the background. Moreover one avoids the complications due to the decay of $\Sigma$(1385) which imposes a difficulty in the analysis.

\subsubsection{Studies with a carbon target inside KLOE}

In 2012 a special setup pure carbon target was inserted into KLOE between the beam pipe and the DC entrance wall (see fig.\label{f3}). In this way we could obtain an almost pure data sample of K$^{-}$ absorption in carbon with an increase in the  statistics of stopped K$^{-}$ at rest.
The sample of pure carbon stops complement the data of the KLOE runs with in-flight absorptions performed till 2005.

\begin{figure}[tbh]
\includegraphics [width= 150mm]{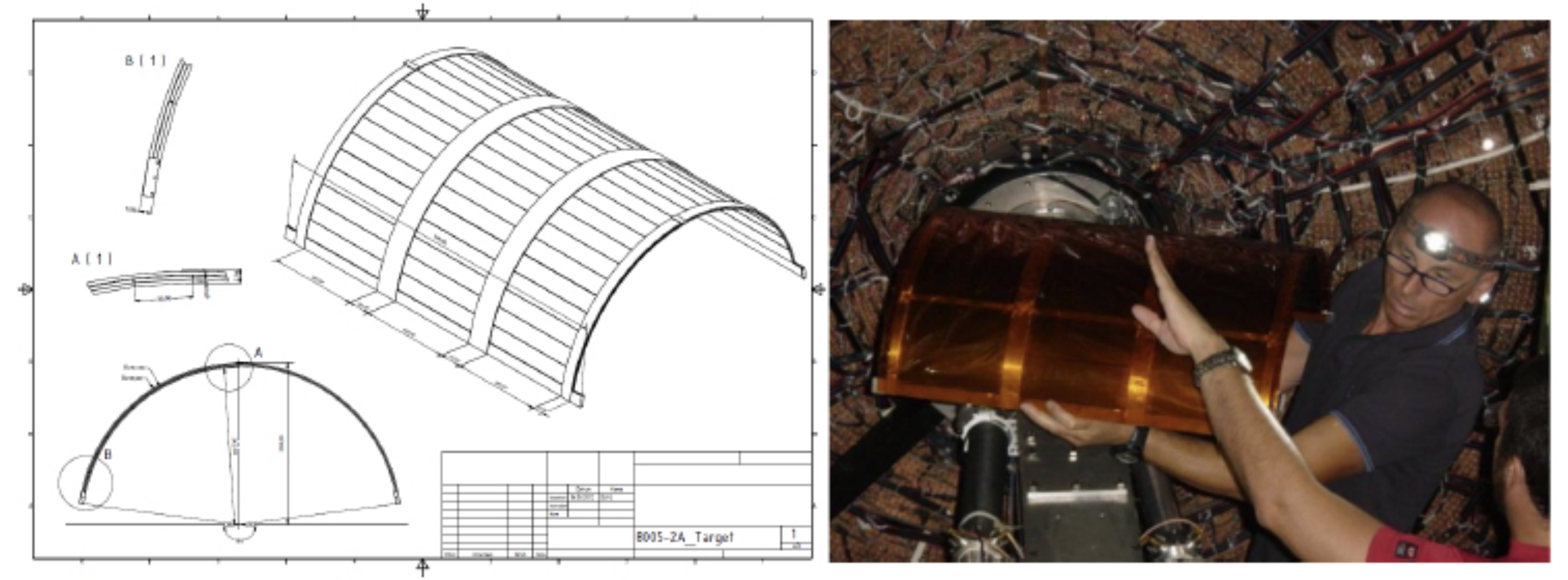}
\caption{For a special run sequence a pure carbon target was inserted in the KLOE apparatus. The stopping efficiency was increased and data on antikaon interaction on carbon nuclei were collected.}
\label{f3}
\end{figure}

\subsection{Outlook: The dedicated AMADEUS Setup}
The dedicated AMADEUS setup aims to investigate the K$^{-}$ interaction with nucleons and light nuclei both at rest and in flight (for K$^{-}$ momenta of about 100 MeV). The goal is to solve longstanding open issues in the nonperturbative QCD in the
strangeness sector, like the nature of the $\Lambda$(1405) state, the resonant versus nonresonant yield in nuclear K$^{-}$ capture
and the properties of possible kaonic nuclear bound states which are strongly
related to the multi-nucleon absorption processes. A measurement of low-momenta kaon
scattering cross-sections in the unexplored momentum region (below 100 MeV/c to threshold) may provide new valuable input for effective field theories used to
describe the kaon nucleon interaction at lowest energies.




\section{Acknowledgement}

For the support in data analysis we like to thank the KLOE collaboration.
We acknowledge the Croatian Science Foundation under Project No. 1680.
Part of this work was supported by the European Community-Research Infrastructure Integrating
Activity “Study of Strongly Interacting Matter” (HadronPhysics2, Grant Agreement No. 227431, and
HadronPhysics3 (HP3) Contract No. 283286) under the EU Seventh Framework Programme.


\begin{thebibliography}{99}

\bibitem{siddharta12} M. Bazzi et al., Phys. Lett. B\textbf{704} (2011) 113, M. Bazzi et al, Nucl. Phys. A\textbf{881} (2012) 88, J. Marton et al. (SIDDHARTA) EPJ Web Conf. \textbf{113} (2016) 03009, arXiv:1603.08755 [nucl-ex].
\bibitem{weise15}  W. Weise, Hyperfine Int. \textbf{233} (2015) 131.
\bibitem{kaiser95} N. Kaiser, B. P. Siegel, W. Weise, Nucl.Phys. A \textbf{594} (1995) 325.
\bibitem{oset98} E. Oset, A. Ramos, Nucl. Phys. A\textbf{635} (1998) 99.
\bibitem{oller01} J. A. Oller, U. G. Mei{\ss}ner, Phys.Lett.B\textbf{500} (2001) 263.
\bibitem{ikeda12} Y. Ikeda, T. Hyodo,W. Weise; Nucl.Phys. A\textbf{881} (2012) 98.
\bibitem{e15} M. Iwasaki et al., Proposal of J-PARC 50-GeV PS “A search for deeply-bound kaonic nuclear states by in-flight 3He(K-,n) reaction”, (2006); Y. Sada et al. (JPARC E15 Coll.) arXiv:1601.06876 [nucl-ex] (2016).
\bibitem{e27} Y. Ichikawa et al., Few-Body Syst. \textbf{54} (2013) 1191.
\bibitem{hyodo12} T. Hyodo, D. Jido, Prog.Part.Nucl.Phys. \textbf{67} (2012) 55.
\bibitem{laura}L. Fabbietti et al.(HADES), Nucl. Phys. A\textbf{835} 333.
\bibitem{moriya13} K. Moriya et al. (CLAS Collaboration) Phys:Rev. C\textbf{87} (2013) 035206.
\bibitem{LOI2006} AMADEUS Letter of Intent, www.lnf.infn.it/esperimenti/siddharta/LOI.
\bibitem{AMADEUS} The AMADEUS collaboration, LNF preprint, LNF/9607/24(IR) (2007).
\bibitem{adinolfi02} M. Adinolfi et al., [KLOE Collaboration], Nucl.Inst. Meth. A\textbf{488} (2002) 51.
\bibitem{adinolfi02a} M. Adinolfi et al. [KLOE Collaboration], Nucl. Inst. Meth. A\textbf{482} (2002) 364.
\bibitem{AY2002} Y. Akaishi and T. Yamazaki, Phys. Rev. C\textbf{65} (2002) 044005.
\bibitem{suzuki2012} K. Suzuki et al., Nucl. Phys. A\textbf{827} (2012) 312C.
\bibitem{agnello05} M. Agnello et al.,Phys.Rev.Lett. \textbf{94} (2005) 919303.
\bibitem{tsuzuki08} T. Suzuki et al., Mod.Phys.Lett.\textbf{A23} (2008) 2520.
\bibitem{catalina14} C. Curceanu et al., Acta Physica Polonica B\textbf{45} (2014) 753.
\bibitem{Kristian} K. Piscicchia et al. e-Print: arXiv:1304.7165.
\bibitem{vazquez12} O. Vazquez Doce et al., Hyperfine Interactions Volume \textbf{210}, Numbers 1-3 (2012).
\bibitem{oton16} O. V\'{a}zquez Doce et al., Phys. Lett B\textbf{758} (2016) 134.


\end{thebibliography}
\end{document}